\documentclass[apj]{emulateapj}
\usepackage[colorlinks,linkcolor={blue},citecolor={blue},urlcolor={red}]{hyperref}
\usepackage{bm}
\usepackage{graphicx}
\usepackage{epsf}
\usepackage{graphics}
\usepackage{amsmath}

\def\beq{\begin{equation}}
\def\eeq{\end{equation}}
\def\bey{\begin{eqnarray}}
\def\eey{\end{eqnarray}}

\def\msun{\, h^{-1}{\rm M_\odot}}
\def\msunt{\, h^{-2}{\rm M_\odot}}

\def\gs{\mathrel{\raise1.16pt\hbox{$>$}\kern-7.0pt
\lower3.06pt\hbox{{$\scriptstyle \sim$}}}}
\def\ls{\mathrel{\raise1.16pt\hbox{$<$}\kern-7.0pt
\lower3.06pt\hbox{{$\scriptstyle \sim$}}}}
\def\gtsima{\, {\buildrel > \over \sim} \,}
\def\ltsima{\, {\buildrel < \over \sim} \,}
\def\prosima{\, {\buildrel \propto \over \sim} \,}
\def\gsim{\lower.5ex\hbox{\gtsima}}
\def\lsim{\lower.5ex\hbox{\ltsima}}
\def\simgt{\lower.5ex\hbox{\gtsima}}
\def\simlt{\lower.5ex\hbox{\ltsima}}
\def\simpr{\lower.5ex\hbox{\prosima}}

\shorttitle{Characteristic masses in quenching}
\shortauthors{P.F. Li et al.}

\begin{document}
%\maketitle
\title{Characteristic mass in galaxy quenching:
environmental versus internal effects}

\author{Pengfei Li\altaffilmark{1,2}, Huiyuan Wang\altaffilmark{1,2}, H.J. Mo\altaffilmark{3}, Enci Wang\altaffilmark{5}, Hui Hong\altaffilmark{1,2}}
\altaffiltext{1}{CAS Key Laboratory for Research in Galaxies and Cosmology, Department of Astronomy, University of Science and Technology of China, Hefei, Anhui 230026, China;lpfv@mail.ustc.edu.cn,
whywang@ustc.edu.cn}
\altaffiltext{2}{School of Astronomy and Space Science, University of Science and Technology of China, Hefei 230026, China}
\altaffiltext{3}{Department of Astronomy, University of Massachusetts, Amherst MA 01003-9305, USA}
%\altaffiltext{4}{Tsinghua Center of Astrophysics \& Department of Physics, Tsinghua University, Beijing 100084, China}
\altaffiltext{5}{Department of Physics, ETH Zurich, Wolfgang-Pauli-Strasse 27, CH-8093 Zurich, Switzerland}
%\date{2019.9.26}
\begin{abstract}
A clear transition feature of galaxy quenching is identified in the multi-parameter 
space of stellar mass ($M_*$), bulge to total mass ratio 
($B/T_{\rm m}$), halo mass ($M_{\rm h}$) and halo-centric distance ($r/r_{180}$).
For given halo mass, the characteristic stellar mass ($M_{*, \rm ch}$)
for the transition is about one-fifth of that of the 
corresponding central galaxy, and almost independent of $B/T_{\rm m}$. 
Once $B/T_{\rm m}$ is fixed, the quenched fraction of galaxies 
with $M_*<M_{*, \rm ch}$ increases with $M_{\rm h}$, but decreases with 
$M_*$ in the inner part of halos ($r/r_{180}<0.5$). 
In the outer part ($r/r_{180}>0.5$), the trend with $M_{\rm h}$ 
remains but the correlation with $M_*$ is absent or becomes positive. 
For galaxies above $M_{\rm *, ch}$ and with $B/T_{\rm m}$ fixed, 
the quenched fraction increases with $M_{\rm *}$, but depends only 
weakly on $M_{\rm h}$ in both the inner and outer regions. 
At fixed $B/T_{\rm m}$ and $M_*$, the quenched fraction increases with 
decreasing $r/r_{180}$ for galaxies with $M_*<M_{*, \rm ch}$, 
and depends only weakly on $r/r_{180}$ for galaxies with $M_*>M_{*, \rm ch}$.
Our finding provides a physically-motivated way to 
classify galaxies in halos into two classes based on their 
quenching properties: an `upper class' with $M_*>M_{\rm *,ch}$ 
and a `lower class' with $M_*<M_{\rm *,ch}$. Environmental quenching is 
important for `lower class' galaxies, while internal quenching 
plays the dominating role for the `upper class'.
\end{abstract}


%%%%%%%%%%%%%%%%%%%%%% SECTION 1%%%%%%%%%%%%%%%%%%%%%%%%
\section{Introduction}
\label{sec_intro}

In the local Universe, galaxies can be divided into two populations according to 
their star formation activity or rest-frame color 
\citep[e.g.][]{Strateva-01,Baldry-04,Brinchmann-04,Wetzel-Tinker-Conroy-12}.
One resides in the star forming main sequence, where 
galaxies in general have blue color and disk-like morphology, 
and the other is the passive or quenched population, in which 
galaxies have red color, spheroid-like morphology,
and little on-going star formation. The quenched population is observed 
to be present at redshift as high as $z=1$ 
\citep[e.g.][]{Bell-04,Ilbert-13, Muzzin-13,Tomczak-14,Barro-17} 
and to grow continuously with time, indicating that quenching 
drives the evolution of the galaxy population over most 
of the Hubble time.

Many mechanisms have been proposed for the quenching of galaxies. 
In broad terms, they can be divided into 
two categories. The first is internal, such as supernova feedback 
\citep[e.g.][]{White-Rees-78, White-Frenk-91,Murray-2011}  
and active galactic nuclei (AGN) feedback 
\citep[e.g.][]{Croton-06,Bower-06,Best-07, Fabian-12, Heckman-2014,He-2019}, 
which may dispel inter-stellar media (ISM) or
prevent gas from cooling, so as to reduce the cold gas reservoir 
for star formation, and morphological quenching \citep[e.g.][]{Martig-09}, 
in which cold gas disk is stablized by a massive bulge. 
The strengths of these mechanisms are expected to depend on 
galaxy stellar mass (or bulge mass) and structural properties, 
so that the fraction of the quenched galaxies (the quenched fraction) 
is expected to correlate with the stellar mass, bulge mass, 
and the bulge to total ratio (hereafter $B/T$) of galaxies.  
This may be the reason why massive, bulge-dominated early type galaxies are 
usually observed to have low on-going star formation activities.
But the explanation of these correlations is not conclusive  
\citep{Lilly-Carollo-16}.

The second category is environmental. The processes in this category include  
ram pressure stripping \citep[e.g.][]{Gunn-Gott-72, Abadi-Moore-Bower-99, WangE-15} 
and tidal stripping \citep[e.g.][]{Toomre-Toomre-72,Read-06}, which  
remove the ISM from galaxies, and strangulation
\citep[e.g.][]{Larson-80, Balogh-Navarro-Morris-00, vandenBosch-08, Weinmann-09,Peng-Maiolino-Cochrane-15}, 
a process that cuts off the replenishment of star forming gas.   
Interaction and merger of a galaxy with other galaxies, 
which can consume or dispel cold gas by triggering star formation 
and/or AGN activity \citep[e.g.][]{Farouki-Shapiro-82,Moore-96, Conselice-Chapman-Windhorst-03, Cox-06, DiMatteo-05}, 
are two other processes in this category. 
These environmental processes are expected to produce a quenched 
fraction that depends on environment, such as the mass of 
the halo that hosts the galaxy, and the distance of the galaxy 
from the center of its host halo (halo-centric distance). 
Moreover, since lower-mass galaxies generally have shallower 
local gravitational potential wells and are more prone to
environmental effects, the trend of quenching with galaxy 
stellar mass produced by environmental processes
is expected to be the opposite to that produced by internal processes. 

With the advent of large photometric and spectroscopic surveys of galaxies, 
the star formation properties of galaxies, 
and their correlation with both internal properties and environment, 
have been investigated extensively 
\citep[e.g.][]{Baldry-06,Weinmann-06,vandenBosch-08,Peng-10,Wetzel-Tinker-Conroy-12,Fang-13,Woo-13,Bluck-14, Knobel-15,Woo-15, WangH-16, Teimoorinia-Bluck-Ellison-16, Liu-19, Bluck-20}.
It is generally believed that the quenching of central galaxies in halos are 
dominated by internal processes, while halo-specific environment plays 
important roles only for satellite galaxies \citep[e.g.][]{vandenBosch-08}.
However, both observations and numerical simulations 
 have suggested that satellites may evolve in the same way as 
 centrals for several Gyrs after they merge into their host halos 
 \citep[e.g.][]{Wetzel-Tinker-Conroy-12, Shi2020}, 
 indicating that internal quenching still dominate 
 during this period of time. Indeed, internal processes may 
 play an important role even when environmental process 
 starts to operate \citep[][]{Bahe-15}. On the other hand, 
 central galaxies may also be affected by environmental effects. 
 For example, such effects have been invoked to explain the 
 similarities between the centrals of splashback halos 
 and satellites \citep{Hirschmann-14}. The observed 
 `galactic conformity' within halos \citep[][]{Weinmann-06}
 and on larger scales \citep{Kauffmann-13} also indicates that centrals 
 and satellites are subjected to some common environmental effects.
 Finally, for massive clusters that contain more than one massive 
 galaxy in the central region, the separation between centrals and 
 satellites is artificial. All these suggest that the simple 
 central-satellite dichotomy 
may be insufficient to account for quenching in the observed galaxy 
population, and that the correlation between quenching and galaxy 
properties is the consequence of both internal and environmental processes.
Disentangling these processes is, therefore, essential to understanding 
galaxy quenching.

Recently, \citet{WangH-18, WangE-18a, WangE-18b,WangE-20} found that galaxies above and 
below one-fifth of the central galaxy stellar mass exhibit different 
dependence of the quenched fraction and galaxy size on the halo-centric distance. 
Based on the relation between central mass and halo mass relation 
proposed in \cite{Yang-Mo-vandenBosch-09},  
\citet{WangE-20} proposed a characteristic stellar mass, 
\begin{equation}\label{eq_mc}
    M_{\rm *,ch}=\frac{M_0}{5} 
    \frac{(M_h/M_1)^{\alpha+\beta}}{(1+M_{\rm h}/M_1)^{\beta}}
\end{equation}
where 
$\log_{10}(M_0/{\rm M}_\odot h^{-2})=10.31$,
$\log_{10}(M_1/{\rm M}_\odot h^{-1})=11.04$,
$\alpha=0.31$, and $\beta=4.54$. 
If such a characteristic mass is real, it will mean that 
galaxy quenching is not driven simply by the central-satellite 
dichotomy, but is determined by the interaction between 
internal and external processes. In this paper, we perform 
a detailed analysis about the significance of the 
characteristic stellar mass in multi-parameter space, 
aiming at distinguishing different quenching processes. 

The paper is organized as follows. 
Section \ref{sec_data} presents galaxy and group catalogs we use, as well as physical 
quantities we derive from them. In Section \ref{sec_fqmc}, we 
investigate how the quenched population is correlated with 
various parameters that quantify internal and external effects, 
focusing on the presence of characteristic stellar mass
scales in the quenching population.
Finally, we summarize our results in Section \ref{sec_sum}.

%%%%%%%%%%%%%%%%%%%%%%%%% SECTION %%%%%%%%%%%%%%%%%%%%%%%%%%%%
\section{Observational data}
\label{sec_data}

Our galaxy sample is taken from the New York University 
Value Added Galaxy Catalog \citep[NYU-VAGC;][]{Blanton-05a}, which 
is based on Sloan Digital Sky Survey (SDSS) DR7 \citep{Abazajian-09}.
Following the construction of the group catalog (see below), 
we select galaxies in the redshift range $z=0.01$-$0.2$, 
with spectroscopic completeness ($C$) 
larger than $0.7$ and with the $r$-band magnitude limit of $17.72$ mag.  
Stellar masses of individual galaxies, $M_{\ast}$ (in unit of $\msunt$), 
are obtained by using the relation between the $(g-r)$ color and 
the stellar mass-to-light ratio assuming a \cite{Kroupa-Weidner-03} 
initial mass function (IMF) \citep{Bell-03}.
Star formation rates (SFRs) of galaxies, adopted from the MPA-JHU 
catalog\footnote{http://www.mpa-garching.mpg.de/SDSS/DR7/},
are estimated from SDSS spectra using an updated version of the 
method presented in \cite{Brinchmann-04} and a Kroupa IMF.
We separate galaxies into a star-forming population and 
a quenched population in the SFR-$M_\ast$ space, as described below.

We use the bulge to total {\it mass} ratio ($B/T_{\rm m}$) of a galaxy 
as an indicator of its morphological type. Extending the work of 
\cite{Simard-11}, \cite{Mendel-14a} obtained the bulge$+$disk 
decomposition for SDSS images in the $u$, $g$, $r$, $i$, and $z$ 
bands. They then estimated the bulge and disk stellar masses 
by modeling the corresponding broadband spectral energy distribution. 
We adopt the ratios between the bulge mass and the 
bulge$+$disk mass given in \cite{Mendel-14a} as the $B/T_{\rm m}$  
for our sample galaxies. We note that we have made tests 
using the bulge-to-total {\it light} ratio in the $r$-band, 
as given in \cite{Simard-11}, instead of $B/T_{\rm m}$, 
and found no significant changes in our results.

\begin{figure*}
    \centering
    \includegraphics[scale=0.45]{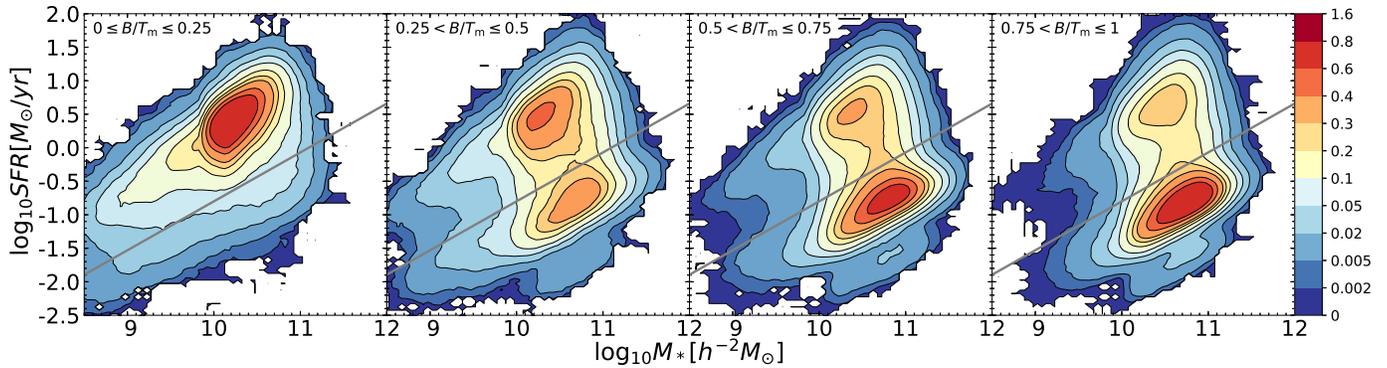}
	\caption{The contours show the
	probability density of galaxies in the $\log (M_\ast)$-$\log ({\rm SFR})$
	space in four $B/T_{\rm m}$ bins as indicated in each panel. The gray line shows the demarcation 
	line proposed by \citet{Bluck-16} to separate star-forming and quenched 
	populations. }
	\label{fig_bimodality}
\end{figure*}

The group catalog used here is constructed by using the halo-based 
group finder of \cite{Yang-07}. The host halo mass ($M_{\rm h}$) of each group is 
estimated using a abundance matching method based on the total 
stellar mass of all member galaxies 
brighter than $M_r = -19.5 + 5\log{h}$ in the $r$-band.
Tests using realistic mock catalogs show that the uncertainty 
in the halo mass estimate is typically 0.2-0.3 dex
\citep{Yang-07, Lim-17}, which is comparable to  
or smaller than the scales we use to smooth the data.
We note that similar smoothing scales are also adopted  
for stellar masses of galaxies to account for potential 
uncertainties in $M_*$.
The most massive galaxy in a group is defined as the central galaxy
of the group while all the rest (if any) as satellite galaxies. 
For each member galaxy in a group, we define a halo-centric distance, 
$r/r_{180}$, where $r$ is the projected distance between the galaxy 
and the luminosity-weighted center of the group. The halo virial radius, 
$r_{180}$, is estimated using equation (5) in \cite{Yang-07}. 
Modeling gravitational lensing effects of the groups 
in \citet{Yang-07}, \citet{Luo-18} found that the  
uncertainties in the group centers are typically within 20\% of 
$r_{180}$. We thus choose $\Delta r/r_{180}=0.2$ to smooth the data
for our analysis.

A total of 23,051 galaxies that do not have estimates of star 
formation rate or bulge-to-total ratio are discarded from our analysis.  
Since we want to study the dependence of quenching on halo environment, 
we only use galaxies for which host halo masses are available from 
\cite{Yang-07}. The lowest halo mass is about $\log( M_{\rm h}/\msun)\sim 11.6$. 
We therefore only select galaxies with $\log( M_{\rm h}/\msun)\geq 11.6$. 
This eliminates 94,504 low-mass galaxies, and our final sample contains 
426,521 galaxies. 

Figure \ref{fig_bimodality} shows the distributions 
in the SFR-$M_*$ space for our sample galaxies in four $B/T_{\rm m}$ bins. 
The distribution is clearly bimodal for the three high $B/T_{\rm }$ 
bins,  and the demarcation line proposed by \cite{Bluck-16} 
can be used to separate the star-forming and quenched populations.
This separation is reliable as long as the demarcation line 
does not cut into one of the modes significantly.   
For the lowest $B/T_{\rm m}$ bin, however, the bimodality is 
rather weak, and so the definition of a quenched fraction 
is uncertain \citep[see also][]{Morselli2017}. We use the same 
demarcation line for the lowest $B/T_{\rm m}$ bin, but keep in 
mind this uncertainty.

%%%%%%%%%%%%%%%%%% SECTION 3 %%%%%%%%%%%%%%%%%%%%%%%%%%%%%
\section{The characteristic stellar mass}
\label{sec_fqmc}

The statistical quantity investigated here is the quenched 
fraction, defined as the fraction of galaxies that are quenched 
according to the criterion described above.  
For a given sub-sample of $N$ galaxies, the quenched fraction, $f_{\rm Q}$, 
is calculated as:
\begin{equation}\label{eq_sts}
    f_{\rm Q}=\frac{\sum^N_{i=1} \omega_i q_i}{\sum^N_{i=1} \omega_{i}}\,,
\end{equation}
where $q_i$ is the quenching property for the $i$th galaxy and 
$\omega_{i}$ is a weight assigned to the galaxy. 
If a galaxy is quenched,  $q_i=1$, otherwise $q_i=0$. The weight is defined 
as  $\omega=1/(V_{\rm max}C)$. Here $V_{\rm max}$, calculated using the 
$K$-correction utilities, v4\_2,  of \cite{Blanton-Roweis-07}, 
is used to correct for the Malmquist bias, and $C$ is the redshift 
incompleteness obtained from the NYU-VAGC.

\begin{figure}
    \centering
    \includegraphics[scale=0.4]{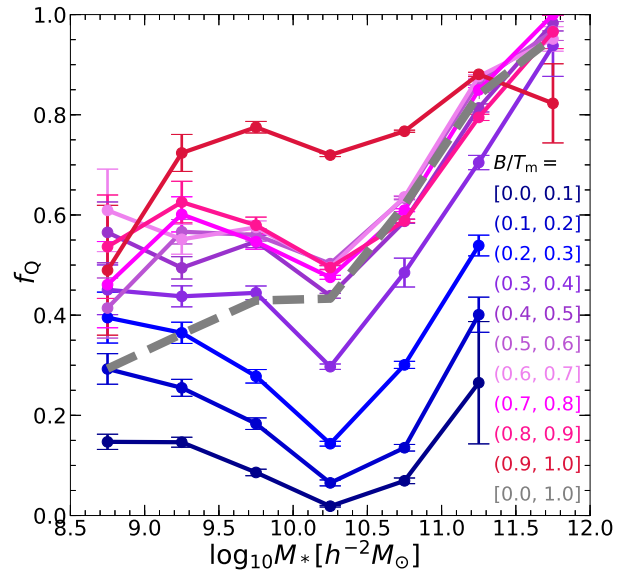}
    \caption{Quenched fraction, $f_{\rm Q}$, as a function of $M_{\rm *}$ for 
    galaxies of different $B/T_{\rm m}$. Here we only use galaxies residing in 
    halos of $\log(M_{\rm h}/\msun)\geq 11.6$, the mass limit of the group 
    sample. Different colors represent different $B/T_{\rm m}$. 
    The grey dashed line shows the result for the whole galaxy sample 
    without separations according to $B/T_{\rm m}$.
    Bins containing less than 10 galaxies are not used. Error bars 
    are estimated by using 1,000 bootstrap samples. }\label{fq_Mstellar_r_BTm_Mh_11o4}
\end{figure}

\begin{figure*}
    \includegraphics[scale=0.46]{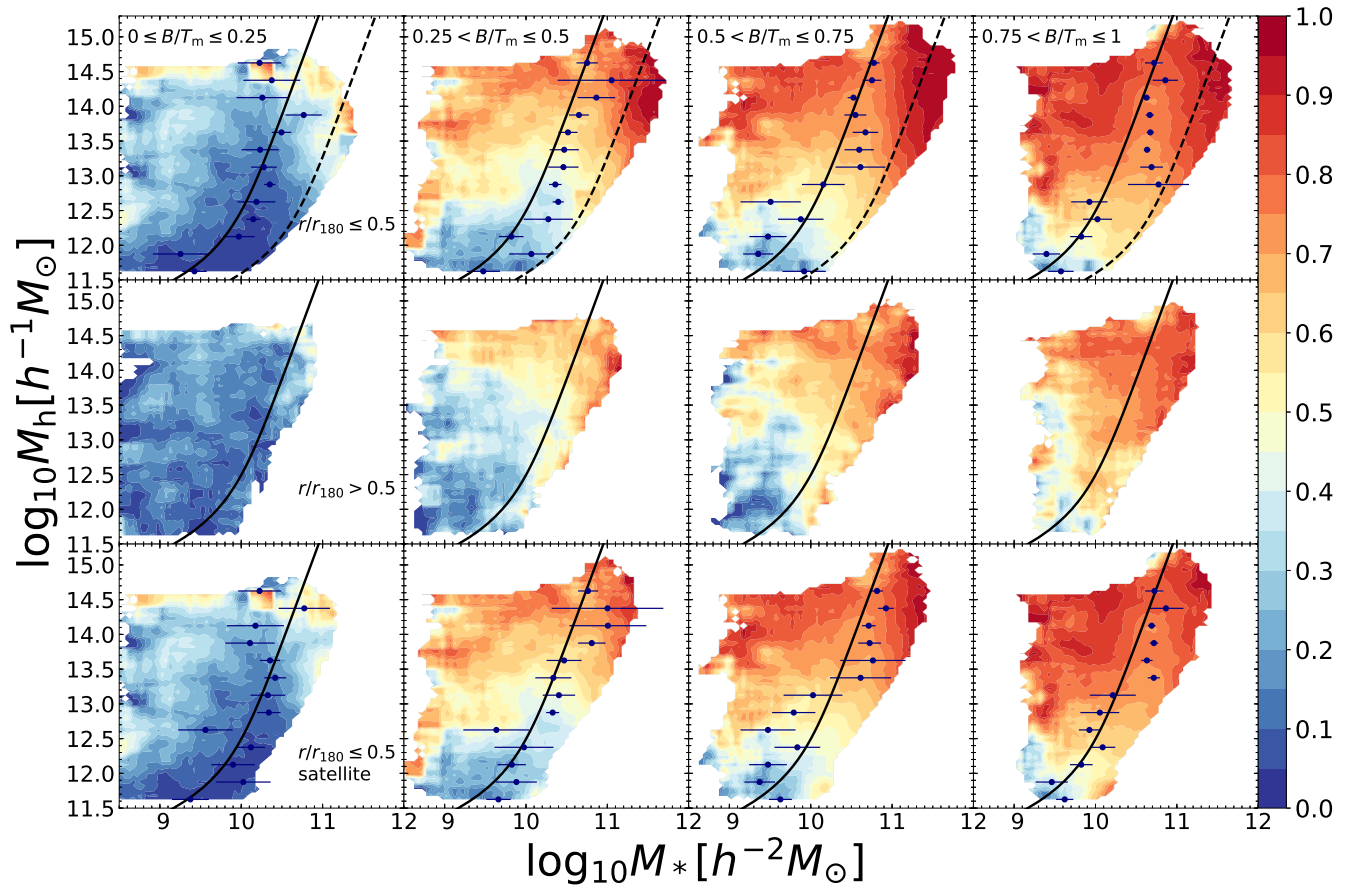}
	\caption{Contours show $f_{\rm Q}$ as a function of $M_{\rm h}$ and $M_{*}$. 
	The top panels are for galaxies with $r/r_{180} \leq 0.5$, the middle panels for 
	$r/r_{180}>0.5$, and the bottom panels for satellites 
	with $r/r_{180} \leq 0.5$. Different columns show
	different $B/T_{\rm m}$, as indicated in the top panels. 
	The results shown are smoothed on a grid with cell size 
	given by $\Delta \log(M_*/\msunt)=0.3$ and 
	$\Delta \log(M_{\rm h}/\msun)=0.3$, but our results are insensitive 
	to changes in the smoothing cell size. 
	Cells with less than 10 galaxies are discarded. Solid dots with 
	error bars show the best-fitting $M_{\rm *,ch}$ and the black solid line shows 
	equation (\ref{eq_mc}). The dashed line in each of the four top panels 
	shows the central mass versus halo mass relation 
	given in \cite{Yang-Mo-vandenBosch-09}.}
	\label{fq_Mh_Mstellar_BTm_r_contour}
\end{figure*}

To start with we first examine how the quenched fraction depends 
on the two intrinsic properties of galaxies, the stellar mass, 
$M_*$ and the bulge-to-total ratio, $B/T_m$. 
Figure \ref{fq_Mstellar_r_BTm_Mh_11o4} shows $f_{\rm Q}$ as a function of $M_{\rm *}$ 
for galaxies with different $B/T_{\rm m}$ residing in halos with
$\log( M_{\rm h}/\msun)\geq11.6$. There is a clear transition in the 
$M_*$-dependence of $f_{\rm Q}$: the quenched fraction first decreases 
with $M_*$ at $\log (M_{\rm *}/\msunt)< 10.3$, and then increases with 
$M_{\rm *}$ at larger masses. This behavior is observed in all $B/T_{\rm m}$ bins. 
\cite{Bluck-14} showed a similar plot in their figure 12, 
but found no transition in the $f_Q$-$M_*$ relation.  
This is because they included galaxies residing in halos with 
$\log(M_{\rm h}/\msun)<11.6$. These galaxies are predominantly
low-mass, star forming galaxies, and including them  
decreases $f_{\rm Q}$ at the low-mass end. 
Furthermore, we found that the transition disappears
if we lump galaxies with different $B/T_{\rm m}$ together,
as shown by the grey dashed line in Figure \ref{fq_Mstellar_r_BTm_Mh_11o4}
(see also the results in \cite{WangH-18} and \cite{WangE-18a}).
This owes to the fact that galaxies of lower masses have, on average, 
smaller $B/T_{\rm m}$. Thus, for the whole sample, the results for 
low-mass and high-mass populations are dominated
by galaxies with small and high $B/T_{\rm m}$, respectively.
The transitional feature seen for galaxies at fixed $B/T_{\rm m}$ 
is erased by the correlation between $B/T_{\rm m}$ and $M_{\rm *}$. 

Clearly, using $M_*$ and $B/T_m$ alone is not sufficient to 
understand the origin of the quenched fraction; to achieve this 
we need to examine how the quenched fraction depends on  
parameters representing both internal and external processes.  
Figure \ref{fq_Mh_Mstellar_BTm_r_contour} shows $f_{\rm Q}$ as a 
function of $M_{\rm h}$ and $M_{\rm *}$ for galaxies 
of different $B/T_{\rm m}$ located both in the 
inner ($r/r_{180} \leq 0.5$) and outer ($r/r_{180} > 0.5$)
regions of halos. We note that, without separating galaxies 
according to $r/r_{180}$, the results are very similar to those 
for $r/r_{180} \leq 0.5$, and that the choice of $r/r_{180}=0.5$
for the separation of inner and outer regions is arbitrary.
A valley-like structure in the $f_{\rm Q}$ field across   
the $M_{\rm h}$-$M_{\rm *}$ plane can clearly be seen for galaxies 
in the inner region and for almost all samples of $B/T_{\rm m}$. 
To the left of the valley, the value of $f_{\rm Q}$ exhibits a negative 
correlation with $M_{\rm *}$ but a positive correlation with 
$M_{\rm h}$. To the right of the valley, in contrast,  
$f_{\rm Q}$ shows a strong, positive correlation with $M_{\rm *}$ 
but only weak dependence on $M_{\rm h}$. The transition feature 
shown in Figure \ref{fq_Mstellar_r_BTm_Mh_11o4} is clearly seen 
here for halos of different masses. This suggests a systematic 
change of the importance of environmental versus internal effects 
as galaxies move across the valley, as we will discuss later. 

The stellar mass at the transition depends on $M_{\rm h}$,  
suggesting the existence of a characteristic 
stellar mass in quenching that is related to halo mass. 
We plot in the upper panels of Figure \ref{fq_Mh_Mstellar_BTm_r_contour} 
the demarcation line (equation \ref{eq_mc}) proposed by 
\citet{WangE-20} based on the relation of galaxy size 
with $r/r_{180}$. Remarkably, the curve lies close to
the bottom of the valley-like structure. To quantify this, 
we measure $M_{\rm *,ch}$ as a function of $M_{\rm h}$ directly 
from our data. The results in Figure \ref{fq_Mstellar_r_BTm_Mh_11o4} suggest 
that $f_{\rm Q}$ as a function of $M_*$ can be approximated by a broken 
power law. We thus use the following function to model the relation: 
\begin{equation}\label{eq_fitfq}
   f_{\rm Q}=\left\{
\begin{aligned}
 a\log M_*/M_{\rm *, ch}(M_{\rm h}) +d \,,(M_*<M_{\rm *,ch})\\
 b\log M_*/M_{\rm *, ch}(M_{\rm h}) +d \,,(M_*\geq M_{\rm *,ch})\,,
\end{aligned}
\right.  
\end{equation}
where $a$, $b$, $d$ and $M_{\rm *,ch}$ are free parameters to be determined. 
The least square method is adopted to perform the fitting and 
the best-fitting $M_{\rm *,ch}$ are shown in 
Figure \ref{fq_Mh_Mstellar_BTm_r_contour}, with 
error bars estimated using 1000 bootstrap samples. 

\begin{figure*}
    \centering
    \includegraphics[scale=0.46]{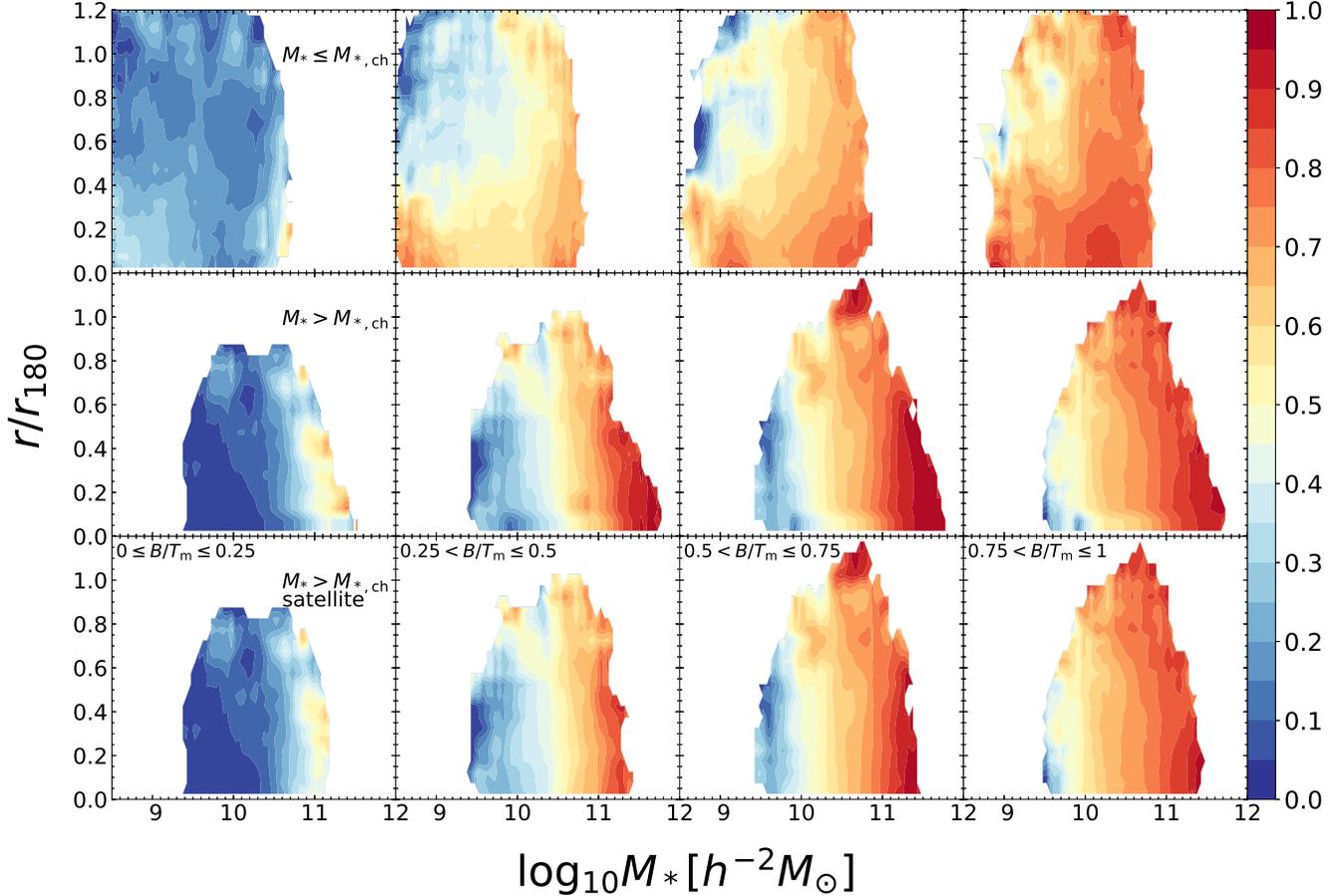}
	\caption{$f_{\rm Q}$ as a function of $r/r_{180}$ and $M_*$. The top panels 
	show galaxies with $M_*<M_{\rm *,ch}$, the middle panels show 
	galaxies with $M_*>M_{\rm *,ch}$ and the bottom panels show satellites with 
	$M_*>M_{\rm *,ch}$. Different columns correspond to different 
	$B/T_{\rm m}$. The grid cell size used in the plot 
	is given by $\Delta \log(M_*/\msunt)=0.3$ and $\Delta r/r_{180}=0.2$;
	the results are insensitive to changes in the cell size.  
	Cell containing less than 10 galaxies are discarded. 
	}\label{fq_r_Mstellar_BTm_Mtstar_contour}
\end{figure*}

\begin{figure*}
    \centering
    \includegraphics[scale=0.5]{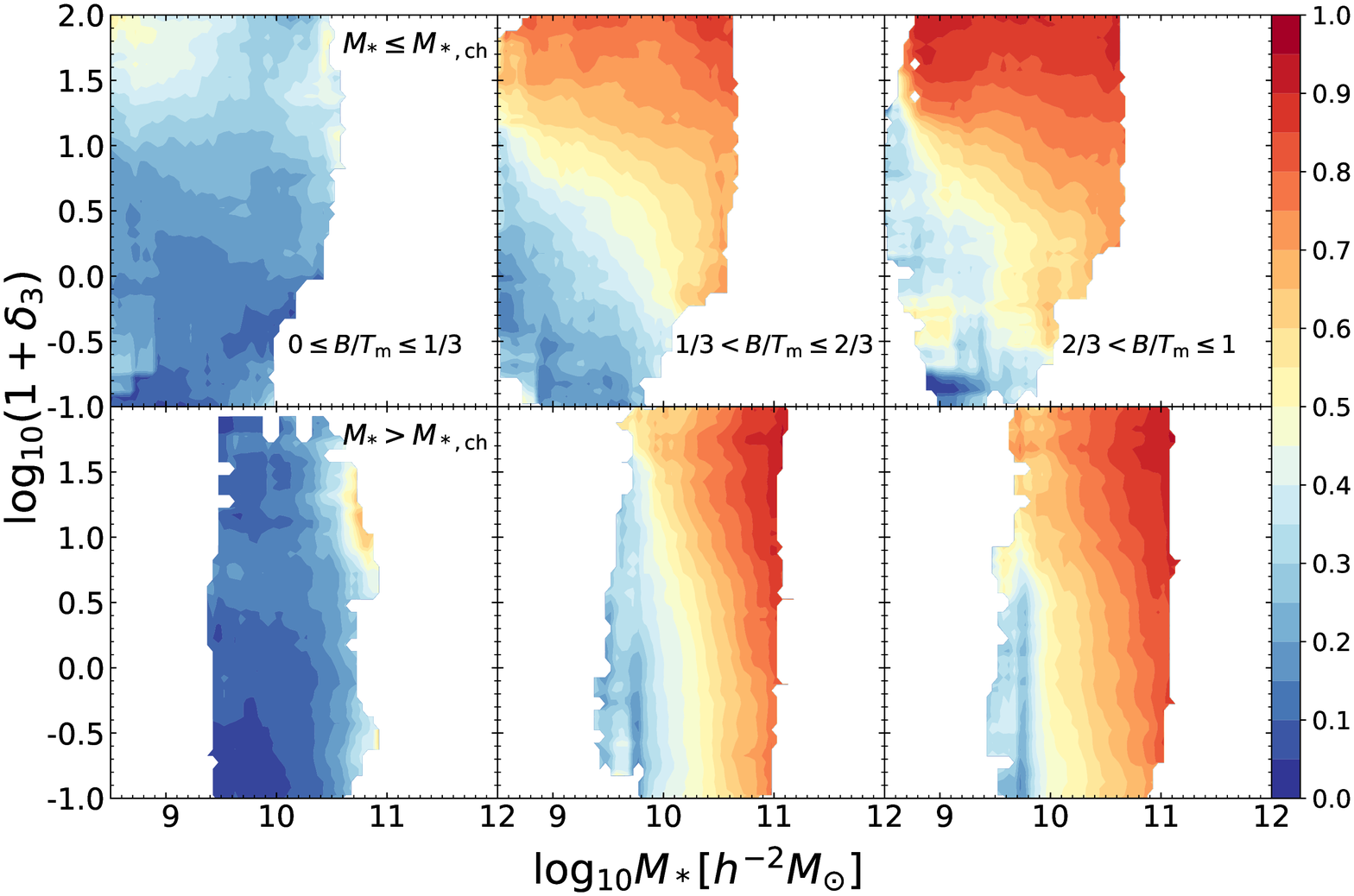}
	\caption{Contours show $f_{\rm Q}$ as a function of $\log(1+\delta_3)$ and $M_*$. 
	Upper panels show results for galaxies in the upper class, i.e.
	with $M_*\le M_{\rm *,ch}$, while lower panels show results  
	for $M_*>M_{\rm *,ch}$. Different columns correspond to different 
	$B/T_{\rm m}$. Results are shown for galaxies with $0.01<z<0.05$. 
	The grid cell size used in the plot 
	is given by $\Delta \log(M_*/\msunt)=0.6$ and $\Delta \log(1+\delta_3)=0.6$;
	the results are insensitive to changes in the cell size.  
	Cells containing less than 10 galaxies are discarded. 
	}\label{fq_NN3_Mstellar_BTm_Mtstar_contour}
\end{figure*}

In general, the best-fitting $M_{\rm *,ch}$ can characterize the 
trend of the valley-like structure, suggesting that the measurement is reliable. 
The measured $M_{\rm *,ch}$-$M_{\rm h}$ 
relation also matches Equation (\ref{eq_mc}) very well, 
except in a small number of bins. This indicates that 
Equation (\ref{eq_mc}) is a good approximation 
for $M_{\rm *,ch}(M_{\rm h})$. 
The fitting results also show that the $M_{\rm *,ch}$-$M_{\rm h}$ 
relation is similar for different $B/T_{\rm m}$. 
This is consistent with the results shown in Figure 
\ref{fq_Mstellar_r_BTm_Mh_11o4}, where the values 
of $M_{\rm *,ch}$ averaged over different $M_{\rm h}$ are  
all close to $10^{10.3}\msunt$ for different $B/T_{\rm m}$. 
However, it is also clear that $M_{\rm *,ch}$ increases with $M_{\rm h}$. 
For Milky-Way size halos, $M_{\rm *, ch}\sim 10^{9.7}\msunt$, 
while for rich clusters, $M_{\rm *, ch}\sim 10^{10.7}\msunt$.

 We also show Equation (\ref{eq_mc}) in the mid-row 
panels of Figure \ref{fq_Mh_Mstellar_BTm_r_contour}. 
Although a valley-like structure is not obvious here, 
galaxies to the left and right of the curve do exhibit different behavior. 
For example, to the left of the curve, the 
correlation with $M_*$ is weak at small $B/T_{\rm m}$ 
and positive at large $B/T_{\rm m}$. To the right of the curve,  
on the other hand, $f_{\rm Q}$ increases with $M_*$ much more 
rapidly. More importantly, the $M_{\rm h}$ dependence is strong 
on the left side of the curve, but is weak or absent on 
the right. 

One possible origin of the transition is the 
central-satellite dichotomy. If centrals behave very differently 
from satellites, a transition feature might be 
produced. We therefore show the central mass - halo mass relation 
obtained by \citep{Yang-Mo-vandenBosch-09} as the dashed lines 
in the upper-row panels to indicate where central galaxies
are likely to be located in the $M_*$-$M_h$ space.
Since the stellar mass - halo mass relation for 
centrals is quite tight and quite far away from 
$M_{\rm *,ch}$-$M_{\rm h}$ relation, one expects that 
the transition feature is dominated by satellite galaxies. 
To demonstrate this directly, the bottom row
of Figure \ref{fq_Mh_Mstellar_BTm_r_contour} shows the 
contour plots of the quenched fraction of satellites 
galaxies, together with the best-fitting $M_{\rm *, ch}$.
The transition feature is also clearly seen for satellites in 
all the $B/T_{\rm m}$ bins, and the best-fitting $M_{\rm *, ch}$ matches 
Equation (\ref{eq_mc}). This indicates that the characteristic 
mass we obtain is not produced by the 
central-satellite dichotomy.

Given the difference in behavior for galaxies below and above $M_{\rm *, ch}$, 
it is interesting to investigate whether or not 
the two populations have different 
dependence on $r/r_{180}$. For NFW halos \citep{NFW97}, 
$r/r_{180}$ is an approximate measure of the local mass density,
independent of halo mass, and so can be used as an indicator 
of local environment within individual halos.
Figure \ref{fq_r_Mstellar_BTm_Mtstar_contour} shows $f_Q$ as a function 
of $r/r_{180}$ and $M_*$ for galaxies of different $B/T_{\rm m}$.  
For galaxies with $M_*\geq M_{\rm *,ch}$, $f_{\rm Q}$ decreases with increasing 
$r/r_{180}$, indicating that such galaxies tend to be more quenched 
in the inner region. In contrast, the $r/r_{180}$ dependence is 
almost absent for $M_*>M_{\rm *,ch}$. The results remain basically 
the same after central galaxies are eliminated, 
as shown in the bottom row of 
Figure \ref{fq_r_Mstellar_BTm_Mtstar_contour}.

Note that we do not distinguish halos of different 
masses in the results shown 
Figure \ref{fq_r_Mstellar_BTm_Mtstar_contour}.
One might think that the results shown above are caused 
by the dependence on halo mass.
However, as shown in \citet{Yang2005b}, the distribution of satellite 
galaxies traces the dark matter profile in a similar way in groups of 
different masses. In fact, \cite{WangH-18} and \cite{WangE-18a} found that, 
for galaxies of given stellar mass, the 
quenched fraction is independent of $r/r_{180}$ for $M_*>M_{\rm *,ch}$ 
and decreases with $r/r_{180}$ for $M_*<M_{\rm *,ch}$, and 
that this is true even when the host halo mass is controlled. 
Our test using galaxies within several halo mass bins,
as shown in Figure \ref{fq_r_Mstellar_BTm_Mh_contour} in 
the Appendix, confirms that the $r/r_{180}$
dependence is not produced by the dependence on halo mass.
Thus, the dependence on $r/r_{180}$ provide independent evidence 
that the characteristic mass, $M_{\rm *,ch}$, defines a 
boundary where environmental effects start to play an 
important role in galaxy quenching. 

%We have also made a test using the nearest neighbours of galaxies to estimate the environmental density \citep[e.g.][]{Baldry-06, Muldrew2012} instead of using $r/r_{180}$. The results, depicted in Fig.\,\ref{fq_NN3_Mstellar_BTm_Mtstar_contour} in the appendix, show again that galaxies below $M_{\rm *, ch}$ exhibit strong environmental dependence while galaxies above $M_{\rm *, ch}$ do not.

 A kink-like structure in the contours of $f_{\rm Q}$ is noticeable 
at the high stellar mass edge for galaxies in upper panels of 
Figure \ref{fq_Mh_Mstellar_BTm_r_contour} and at $r/r_{180}\sim 0$ 
in the middle panels of Figure \ref{fq_r_Mstellar_BTm_Mtstar_contour}. 
Their high stellar mass for their halo mass and their small 
halo-centric distance suggest that these galaxies are dominated by 
centrals. Indeed, the structure disappears if centrals are eliminated.
Unfortunately, it is unclear if this is a real feature, 
or a false signal produced by the halo mass assignment technique based on  
the total stellar mass that is dominated by the most massive 
galaxies for some groups. 

As an independent check of our results 
on the difference between the upper and lower classes, 
we use the commonly-adopted nearest neighbour galaxy density 
\citep[e.g.][]{Baldry-06,Muldrew2012} to represent the local 
environment. To obtain a reliable galaxy density measurement, 
here we only focus on the low redshift galaxy sample, 
with $0.01\leq z\leq 0.05$). We have checked our results 
using $0.01\leq z\leq 0.1$ and reached similar conclusions. 
For each galaxy (target) in the sample, we calculate 
the local density as $\Sigma_3=3/\pi d_3^2$, 
where $d_3$ is the projected distance of the target to its 
third nearest neighbour that belongs to a volume-limited sample 
and has velocity difference less than $1000\,\rm km/s$
with the target. We define a normalized density as
$1+\delta_3=\Sigma_3/\bar\Sigma_3$, where $\bar\Sigma_3$ is 
the median density of all targets in the sample. 
Figure \ref{fq_NN3_Mstellar_BTm_Mtstar_contour} 
shows the quenched fraction as a function of $M_*$ and $1+\delta_3$. 
As one can see, galaxies below $M_{\rm *, ch}$ show strong dependence 
on $1+\delta_3$, while above $M_{\rm *, ch}$ the 
dependence is much weaker. This is consistent with the results 
based on $r/r_{180}$ and $M_{\rm h}$. There is weak but significant 
dependence of $f_{\rm Q}$ on $\delta_3$ for the upper class galaxies, 
which is in qualitative agreement with the weak, positive correlation 
between $f_Q$ and $M_h$ seen in Fig.\,\ref{fq_Mh_Mstellar_BTm_r_contour}.

We conclude this section with some discussion about 
the uncertainties that may affect our results. 
As shown in Figure \ref{fig_bimodality}, galaxies in the lowest 
$B/T_{\rm m}$ bin lack a clear bimodal distribution in the SFR-$M_*$ space. 
This makes the separation between the 
star-forming and quenched populations uncertain. 
However, as shown above, the transition feature is found 
for all the $B/T_{\rm m}$ bins and the characteristic mass 
is almost independent of $B/T_{\rm m}$. This suggests 
that the use of the same demarcation line for low-$B/T_{\rm m}$ galaxies
is reasonable and that the transition feature seen for galaxies 
of different $B/T_{\rm m}$ may have a similar origin.
There are also potential uncertainties introduced by 
effects like fiber aperture, sample completeness
and bulge$+$disk decomposition, which are expected 
to be redshift dependent. As a test, we have made the same 
analysis but using galaxies in smaller redshift ranges, 
$0.01\leq z\leq 0.05$ and $0.01\leq z\leq 0.1$, and the results 
are presented in the Appendix 
(see Fig.\,\ref{fq_Mh_Mstellar_BTm_r_contour_z}).
The transition feature is clearly seen in these results,
albeit noisier than that for the whole sample because of the 
smaller sample size. This demonstrates that the characteristic mass 
we find is not a result of these uncertainties.

\section{Summary and discussion}\label{sec_sum}

In this paper, we have investigated how galaxy quenching in 
star formation depends on internal properties of galaxies, 
represented by their stellar mass and $B/T_{\rm m}$, and on environment, 
represented by the mass of their host halos and 
their distances to the halo centers. Our main results can be summarized as follows.
\begin{itemize}
    \item For a given $B/T_{\rm m}$ and a given host halo mass, 
          there is a characteristic stellar mass, 
          $M_{*, \rm ch}$, where $f_{\rm Q}$ is the lowest. 
          The characteristic mass is about one-fifth of that of the 
          central galaxy of the halo, almost independent of $B/T_{\rm m}$.
          The presence of the characteristic mass is more prominent 
          in the inner region of halos. 
    \item Galaxies with masses below and above $M_{*, \rm ch}$ 
          have different quenching properties.
          At $M_*<M_{*, \rm ch}$, $f_{\rm Q}$ for galaxies of a given 
          $B/T_{\rm m}$ decreases with $M_*$ and $r/r_{180}$ but 
          increases with $M_{\rm h}$. 
          At $M_*\geq M_{*, \rm ch}$, however, $f_{\rm Q}$ increases with $M_*$ but 
          depends only weakly on $M_h$ and $r/r_{180}$. The trend with $M_{\rm h}$ 
          is similar in both the inner and outer regions,
          but the decrease of $f_{\rm Q}$ with $M_*$ at $M_*< M_{*, {\rm ch}}$
          is absent in the outer region. Similar conclusions are reached using 
          $\delta_3$ to represent the environmental density.
    \item The characteristic stellar mass is independent of the 
          central-satellite dichotomy.     
\end{itemize}

Observationally, the most massive/luminous galaxy in a galaxy groups/clusters 
is usually classified as the central, while the rest as satellites.
As mentioned in the introduction, such classification may not be able to 
truly reflect the physical processes operating on them. This definition 
is also different from that in theoretical models and simulations, 
where a central galaxy usually refers to the one located 
near the center of the host halo or the bottom of the 
gravitational potential well. Given that galaxies below and 
above $M_{\rm *,ch}$ have systematically different quenching 
properties, our finding provides an physically-motivated 
dichotomy for galaxies in halos, an `upper class' with $M_*>M_{\rm *,ch}$ 
and a `lower class' with $M_*<M_{\rm *,ch}$.

`Lower class' galaxies exhibit strong dependence of $f_{\rm Q}$ 
on both $M_h$ and $r/r_{180}$ even with both $M_*$ and $B/T_{\rm m}$
controlled (Figs.~\ref{fq_Mh_Mstellar_BTm_r_contour} and \ref{fq_r_Mstellar_BTm_Mtstar_contour}). 
This clearly indicates that quenching of these galaxies 
is strongly affected by environment. Theoretically, many environmental processes, 
such as ram pressure and tidal stripping, predict a higher efficiency 
for galaxies of lower mass, because of their shallower 
gravitational potential wells. The decrease of $f_{\rm Q}$ with $M_*$
at the low-mass end thus provides a strong evidence for such 
environmental quenching.  
The interpretation is not unique though, as the final environmental 
effect depends not only on the strength of quenching mechanisms, 
but also on how long these mechanisms operate.
For example, \citet{Shi2020} found that satellites accreted earlier 
by their host halos are more frequently quenched. 
Thus, if the accretion time for a satellite depends on $M_{\rm h}$, $M_*$ and 
$r/r_{180}$, the observed correlations can also be produced even if the 
strengths of the quenching mechanisms are independent of the three parameters. 
More investigations are needed to disentangle the dependencies 
on accretion time and quenching strength.

One interesting property of the `lower class' is that
the dependence of $f_Q$ on $M_{\rm h}$ and $r/r_{180}$ 
is much stronger for higher $B/T_{\rm m}$
galaxies (Figs.~\ref{fq_Mh_Mstellar_BTm_r_contour} and 
\ref{fq_r_Mstellar_BTm_Mtstar_contour}). This indicates that 
environmental processes are more effective in quenching 
galaxies of higher $B/T_{\rm m}$. One possibility is that 
the ISM in higher $B/T_{\rm m}$ galaxies is more fluffy and easier to 
strip, as shown in hydrodynamic simulations \citep{Bahe-15}.
Another possibility is that galaxies with higher 
$B/T_{\rm m}$ are accreted into their hosts earlier, so that 
the impact by environment is larger. Indeed, galaxies formed 
earlier are more compact \citep{Mo-Mao-White-99, vanderWel-14} 
and more likely to become galaxies with higher 
$B/T_{\rm m}$ at $z=0$. In both possibilities, environmental 
effects may affect the star formation of a galaxy without 
changing its structure. Thus, unless $B/T_{\rm m}$ is created 
with strong dependence on $M_{\rm h}$ and $r/r_{180}$, the 
weak dependence of galaxy morphology on $M_{\rm h}$ and 
$r/r_{180}$ \citep{Liu-19} suggests that galaxy 
morphology may indeed have formed without strong dependence on environment.
Our results strongly suggest that galaxy morphology, as represented 
by $B/T_{\rm m}$, is an important factor in affecting the efficiency
of both internal and environmental quenching processes.

For galaxies in the `upper class', $f_{\rm Q}$ depends only weakly 
on $M_{\rm h}$, $r/r_{180}$ and $\delta_3$, indicating that their quenching is 
dominated by internal processes. The strong increase of $f_{\rm Q}$ with 
both $M_*$ and $B/T_{\rm m}$ (Figs.~\ref{fq_Mstellar_r_BTm_Mh_11o4} 
and \ref{fq_Mh_Mstellar_BTm_r_contour}) 
at the massive end provides support to such interpretation, 
and suggests that the existence of a significant bulge is essential 
for galaxy quenching \citep[e.g.][]{Bluck-14}. A number of quenching 
mechanisms linked to galaxy bulge have been proposed. 
The morphological quenching, proposed in \citet{Martig-09}, is 
expected to predict strong dependence of $f_Q$ on $B/T_{\rm m}$. 
However, it is unclear if the model can predict the strong $M_*$-dependence 
at fixed $B/T_{\rm m}$. Various types of AGN feedback
models assume that the quenching of massive galaxies is produced by the
energy generated by super-massive black holes whose mass is 
strongly correlated with the velocity dispersion/stellar mass 
of the bulge \citep{Kormendy-Ho-13}. These models, therefore, predict a strong 
positive correlation between $f_{\rm Q}$ and the bulge mass, as seen 
in our results. However, there are still uncertainties 
regarding the strength and mode of AGN feedback, and competing 
models have been proposed \citep[see e.g.][]{Heckman-2014}.
Furthermore, star formation associated with the formation of 
the bulge may also suppress subsequent gas accretion and star 
formation in the disk \cite[e.g.][]{MoMao2004}. 
Finally, satellite galaxies in the `upper class' are also subjected 
to environmental effects. However, because of their large 
stellar mass, these galaxies may be able to maintain their gas 
reservoir before merging with the central galaxies, so that 
environmental effects play only a minor role in quenching 
their star formation.

The characteristic mass scale is thus a result of
the competition between internal and environmental processes. 
In a halo of a given mass, both processes are expected to 
have comparable impacts on quenching for galaxies around 
the characteristic mass. Understanding the origin of the 
simple relation between $M_{\rm *,ch}$ and $M_{\rm h}$ obtained 
here is, therefore, crucial for understanding the nature of both 
environmental and internal quenching mechanisms.

\section*{Acknowledgments}
We thank the referee for useful comments that significantly 
improve the paper.
This work is supported by the National Key R\&D Program of China (grant No. 2018YFA0404503), the National Natural Science Foundation of China (NSFC, Nos.  11733004, 11421303, 11890693, and 11522324), the National Basic Research Program of China (973 Program)(2015CB857002), and the Fundamental Research Funds for the Central Universities. P.F.L. is supported by the Fund for Fostering Talents in Basic Science of the National Natural Science Foundation of China NO.J1310021. EW is supported by the Swiss National Science Foundation. The work is also supported by the Supercomputer Center of University of Science and Technology of China. This research utilized: NumPy \citep{numpy}, a fundamental package for scientific computing with Python; SciPy \citep{scipy}, a Python based ecosystem of open software for mathmatics, science and engineering; Matplotlib \citep{Hunter-2007}, a comprehensive library for creating static, animated and interactive visualizations in Python; Astropy \citep{astropy-2018}, a community python library for astronomy.

%\begin{figure*}
%    \centering
%    \includegraphics[scale=0.5]{fq_r_Mstellar_BTm_Mtstar_contour_z.eps}
%	\caption{}
%	\label{fq_r_Mstellar_BTm_Mtstar_contour_z}
%\end{figure*}

\bibliography{rewritebib.bib}

\begin{thebibliography}{80}
\expandafter\ifx\csname natexlab\endcsname\relax\def\natexlab#1{#1}\fi

\bibitem[{{Abadi} {et~al.}(1999){Abadi}, {Moore}, \&
  {Bower}}]{Abadi-Moore-Bower-99}
{Abadi}, M.~G., {Moore}, B., \& {Bower}, R.~G. 1999, \mnras, 308, 947

\bibitem[{{Abazajian} {et~al.}(2009){Abazajian}, {Adelman-McCarthy},
  {Ag{\"u}eros}, {Allam}, {Allende Prieto}, {An}, {Anderson}, {Anderson}, \&
  {et al.}}]{Abazajian-09}
{Abazajian}, K.~N., {Adelman-McCarthy}, J.~K., {Ag{\"u}eros}, M.~A., {et~al.}
  2009, \apjs, 182, 543

\bibitem[{{Bah{\'e}} \& {McCarthy}(2015)}]{Bahe-15}
{Bah{\'e}}, Y.~M., \& {McCarthy}, I.~G. 2015, \mnras, 447, 969

\bibitem[{{Baldry} {et~al.}(2006){Baldry}, {Balogh}, {Bower}, {Glazebrook},
  {Nichol}, {Bamford}, \& {Budavari}}]{Baldry-06}
{Baldry}, I.~K., {Balogh}, M.~L., {Bower}, R.~G., {et~al.} 2006, \mnras, 373,
  469

\bibitem[{{Baldry} {et~al.}(2004){Baldry}, {Glazebrook}, {Brinkmann},
  {Ivezi{\'c}}, {Lupton}, {Nichol}, \& {Szalay}}]{Baldry-04}
{Baldry}, I.~K., {Glazebrook}, K., {Brinkmann}, J., {et~al.} 2004, \apj, 600,
  681

\bibitem[{{Balogh} {et~al.}(2000){Balogh}, {Navarro}, \&
  {Morris}}]{Balogh-Navarro-Morris-00}
{Balogh}, M.~L., {Navarro}, J.~F., \& {Morris}, S.~L. 2000, \apj, 540, 113

\bibitem[{{Barro} {et~al.}(2017){Barro}, {Faber}, {Koo}, {Dekel}, {Fang},
  {Trump}, {P{\'e}rez-Gonz{\'a}lez}, {Pacifici}, \& {et al.}}]{Barro-17}
{Barro}, G., {Faber}, S.~M., {Koo}, D.~C., {et~al.} 2017, \apj, 840, 47

\bibitem[{{Bell} {et~al.}(2003){Bell}, {McIntosh}, {Katz}, \&
  {Weinberg}}]{Bell-03}
{Bell}, E.~F., {McIntosh}, D.~H., {Katz}, N., \& {Weinberg}, M.~D. 2003, \apjs,
  149, 289

\bibitem[{{Bell} {et~al.}(2004){Bell}, {Wolf}, {Meisenheimer}, {Rix}, {Borch},
  {Dye}, {Kleinheinrich}, {Wisotzki}, \& {et al.}}]{Bell-04}
{Bell}, E.~F., {Wolf}, C., {Meisenheimer}, K., {et~al.} 2004, \apj, 608, 752

\bibitem[{{Best} {et~al.}(2007){Best}, {von der Linden}, {Kauffmann},
  {Heckman}, \& {Kaiser}}]{Best-07}
{Best}, P.~N., {von der Linden}, A., {Kauffmann}, G., {Heckman}, T.~M., \&
  {Kaiser}, C.~R. 2007, \mnras, 379, 894

\bibitem[{{Blanton} \& {Roweis}(2007)}]{Blanton-Roweis-07}
{Blanton}, M.~R., \& {Roweis}, S. 2007, \aj, 133, 734

\bibitem[{{Blanton} {et~al.}(2005){Blanton}, {Schlegel}, {Strauss},
  {Brinkmann}, {Finkbeiner}, {Fukugita}, {Gunn}, {Hogg}, \& {et
  al.}}]{Blanton-05a}
{Blanton}, M.~R., {Schlegel}, D.~J., {Strauss}, M.~A., {et~al.} 2005, \aj, 129,
  2562

\bibitem[{{Bluck} {et~al.}(2020){Bluck}, {Maiolino}, {S{\'a}nchez}, {Ellison},
  {Thorp}, {Piotrowska}, {Teimoorinia}, \& {Bundy}}]{Bluck-20}
{Bluck}, A. F.~L., {Maiolino}, R., {S{\'a}nchez}, S.~F., {et~al.} 2020, \mnras,
  492, 96

\bibitem[{{Bluck} {et~al.}(2014){Bluck}, {Mendel}, {Ellison}, {Moreno},
  {Simard}, {Patton}, \& {Starkenburg}}]{Bluck-14}
{Bluck}, A.~F.~L., {Mendel}, J.~T., {Ellison}, S.~L., {et~al.} 2014, \mnras,
  441, 599

\bibitem[{{Bluck} {et~al.}(2016){Bluck}, {Mendel}, {Ellison}, {Patton},
  {Simard}, {Henriques}, {Torrey}, {Teimoorinia}, \& {et al.}}]{Bluck-16}
---. 2016, \mnras, 462, 2559

\bibitem[{{Bower} {et~al.}(2006){Bower}, {Benson}, {Malbon}, {Helly}, {Frenk},
  {Baugh}, {Cole}, \& {Lacey}}]{Bower-06}
{Bower}, R.~G., {Benson}, A.~J., {Malbon}, R., {et~al.} 2006, \mnras, 370, 645

\bibitem[{{Brinchmann} {et~al.}(2004){Brinchmann}, {Charlot}, {White},
  {Tremonti}, {Kauffmann}, {Heckman}, \& {Brinkmann}}]{Brinchmann-04}
{Brinchmann}, J., {Charlot}, S., {White}, S.~D.~M., {et~al.} 2004, \mnras, 351,
  1151

\bibitem[{{Conselice} {et~al.}(2003){Conselice}, {Chapman}, \&
  {Windhorst}}]{Conselice-Chapman-Windhorst-03}
{Conselice}, C.~J., {Chapman}, S.~C., \& {Windhorst}, R.~A. 2003, \apjl, 596,
  L5

\bibitem[{{Cox} {et~al.}(2006){Cox}, {Jonsson}, {Primack}, \&
  {Somerville}}]{Cox-06}
{Cox}, T.~J., {Jonsson}, P., {Primack}, J.~R., \& {Somerville}, R.~S. 2006,
  \mnras, 373, 1013

\bibitem[{{Croton} {et~al.}(2006){Croton}, {Springel}, {White}, {De Lucia},
  {Frenk}, {Gao}, {Jenkins}, {Kauffmann}, \& {et al.}}]{Croton-06}
{Croton}, D.~J., {Springel}, V., {White}, S.~D.~M., {et~al.} 2006, \mnras, 365,
  11

\bibitem[{{Di Matteo} {et~al.}(2005){Di Matteo}, {Springel}, \&
  {Hernquist}}]{DiMatteo-05}
{Di Matteo}, T., {Springel}, V., \& {Hernquist}, L. 2005, \nat, 433, 604

\bibitem[{{Fabian}(2012)}]{Fabian-12}
{Fabian}, A.~C. 2012, \araa, 50, 455

\bibitem[{{Fang} {et~al.}(2013){Fang}, {Faber}, {Koo}, \& {Dekel}}]{Fang-13}
{Fang}, J.~J., {Faber}, S.~M., {Koo}, D.~C., \& {Dekel}, A. 2013, \apj, 776, 63

\bibitem[{{Farouki} \& {Shapiro}(1982)}]{Farouki-Shapiro-82}
{Farouki}, R.~T., \& {Shapiro}, S.~L. 1982, \apj, 259, 103

\bibitem[{{Gunn} \& {Gott}(1972)}]{Gunn-Gott-72}
{Gunn}, J.~E., \& {Gott}, III, J.~R. 1972, \apj, 176, 1

\bibitem[{{He} {et~al.}(2019){He}, {Wang}, {Liu}, {Wang}, {Bian},
  {Tchernyshyov}, {Mou}, {Xu}, {Zhou}, {Green}, \& {Xu}}]{He-2019}
{He}, Z., {Wang}, T., {Liu}, G., {et~al.} 2019, Nature Astronomy, 3, 265

\bibitem[{{Heckman} \& {Best}(2014)}]{Heckman-2014}
{Heckman}, T.~M., \& {Best}, P.~N. 2014, \araa, 52, 589

\bibitem[{{Hirschmann} {et~al.}(2014){Hirschmann}, {De Lucia}, {Wilman},
  {Weinmann}, {Iovino}, {Cucciati}, {Zibetti}, \& {Villalobos}}]{Hirschmann-14}
{Hirschmann}, M., {De Lucia}, G., {Wilman}, D., {et~al.} 2014, \mnras, 444,
  2938

\bibitem[{Hunter(2007)}]{Hunter-2007}
Hunter, J.~D. 2007, Computing in Science \& Engineering, 9, 90

\bibitem[{{Ilbert} {et~al.}(2013){Ilbert}, {McCracken}, {Le F{\`e}vre},
  {Capak}, {Dunlop}, {Karim}, {Renzini}, {Caputi}, {Boissier}, {Arnouts},
  {Aussel}, {Comparat}, {Guo}, {Hudelot}, {Kartaltepe}, {Kneib}, {Krogager},
  {Le Floc'h}, {Lilly}, {Mellier}, {Milvang-Jensen}, {Moutard}, {Onodera},
  {Richard}, {Salvato}, {Sanders}, {Scoville}, {Silverman}, {Taniguchi},
  {Tasca}, {Thomas}, {Toft}, {Tresse}, {Vergani}, {Wolk}, \&
  {Zirm}}]{Ilbert-13}
{Ilbert}, O., {McCracken}, H.~J., {Le F{\`e}vre}, O., {et~al.} 2013, \aap, 556,
  A55

\bibitem[{Jones {et~al.}(2001--)Jones, Oliphant, Peterson, {et~al.}}]{scipy}
Jones, E., Oliphant, T., Peterson, P., {et~al.} 2001--, {SciPy}: Open source
  scientific tools for {Python}

\bibitem[{{Kauffmann} {et~al.}(2013){Kauffmann}, {Li}, {Zhang}, \&
  {Weinmann}}]{Kauffmann-13}
{Kauffmann}, G., {Li}, C., {Zhang}, W., \& {Weinmann}, S. 2013, \mnras, 430,
  1447

\bibitem[{{Knobel} {et~al.}(2015){Knobel}, {Lilly}, {Woo}, \& {Kova{\v
  c}}}]{Knobel-15}
{Knobel}, C., {Lilly}, S.~J., {Woo}, J., \& {Kova{\v c}}, K. 2015, \apj, 800,
  24

\bibitem[{{Kormendy} \& {Ho}(2013)}]{Kormendy-Ho-13}
{Kormendy}, J., \& {Ho}, L.~C. 2013, \araa, 51, 511

\bibitem[{{Kroupa} \& {Weidner}(2003)}]{Kroupa-Weidner-03}
{Kroupa}, P., \& {Weidner}, C. 2003, \apj, 598, 1076

\bibitem[{{Larson} {et~al.}(1980){Larson}, {Tinsley}, \&
  {Caldwell}}]{Larson-80}
{Larson}, R.~B., {Tinsley}, B.~M., \& {Caldwell}, C.~N. 1980, \apj, 237, 692

\bibitem[{{Lilly} \& {Carollo}(2016)}]{Lilly-Carollo-16}
{Lilly}, S.~J., \& {Carollo}, C.~M. 2016, \apj, 833, 1

\bibitem[{{Lim} {et~al.}(2017){Lim}, {Mo}, {Lu}, {Wang}, \& {Yang}}]{Lim-17}
{Lim}, S.~H., {Mo}, H.~J., {Lu}, Y., {Wang}, H., \& {Yang}, X. 2017, \mnras,
  470, 2982

\bibitem[{{Liu} {et~al.}(2019){Liu}, {Hao}, {Wang}, \& {Yang}}]{Liu-19}
{Liu}, C., {Hao}, L., {Wang}, H., \& {Yang}, X. 2019, \apj, 878, 69

\bibitem[{{Luo} {et~al.}(2018){Luo}, {Yang}, {Lu}, {Shi}, {Zhang}, {Mo}, {Shu},
  {Fu}, {Radovich}, {Zhang}, {Li}, {Sunayama}, \& {Wang}}]{Luo-18}
{Luo}, W., {Yang}, X., {Lu}, T., {et~al.} 2018, \apj, 862, 4

\bibitem[{{Martig} {et~al.}(2009){Martig}, {Bournaud}, {Teyssier}, \&
  {Dekel}}]{Martig-09}
{Martig}, M., {Bournaud}, F., {Teyssier}, R., \& {Dekel}, A. 2009, \apj, 707,
  250

\bibitem[{{Mendel} {et~al.}(2014){Mendel}, {Simard}, {Palmer}, {Ellison}, \&
  {Patton}}]{Mendel-14a}
{Mendel}, J.~T., {Simard}, L., {Palmer}, M., {Ellison}, S.~L., \& {Patton},
  D.~R. 2014, \apjs, 210, 3

\bibitem[{{Mo} \& {Mao}(2004)}]{MoMao2004}
{Mo}, H.~J., \& {Mao}, S. 2004, \mnras, 353, 829

\bibitem[{{Mo} {et~al.}(1999){Mo}, {Mao}, \& {White}}]{Mo-Mao-White-99}
{Mo}, H.~J., {Mao}, S., \& {White}, S.~D.~M. 1999, \mnras, 304, 175

\bibitem[{{Moore} {et~al.}(1996){Moore}, {Katz}, {Lake}, {Dressler}, \&
  {Oemler}}]{Moore-96}
{Moore}, B., {Katz}, N., {Lake}, G., {Dressler}, A., \& {Oemler}, A. 1996,
  \nat, 379, 613

\bibitem[{{Morselli} {et~al.}(2017){Morselli}, {Popesso}, {Erfanianfar}, \&
  {Concas}}]{Morselli2017}
{Morselli}, L., {Popesso}, P., {Erfanianfar}, G., \& {Concas}, A. 2017, \aap,
  597, A97

\bibitem[{{Muldrew} {et~al.}(2012){Muldrew}, {Croton}, {Skibba}, {Pearce},
  {Ann}, {Baldry}, {Brough}, {Choi}, {Conselice}, {Cowan}, {Gallazzi}, {Gray},
  {Gr{\"u}tzbauch}, {Li}, {Park}, {Pilipenko}, {Podgorzec}, {Robotham},
  {Wilman}, {Yang}, {Zhang}, \& {Zibetti}}]{Muldrew2012}
{Muldrew}, S.~I., {Croton}, D.~J., {Skibba}, R.~A., {et~al.} 2012, \mnras, 419,
  2670

\bibitem[{{Murray} {et~al.}(2011){Murray}, {M{\'e}nard}, \&
  {Thompson}}]{Murray-2011}
{Murray}, N., {M{\'e}nard}, B., \& {Thompson}, T.~A. 2011, \apj, 735, 66

\bibitem[{{Muzzin} {et~al.}(2013){Muzzin}, {Marchesini}, {Stefanon}, {Franx},
  {McCracken}, {Milvang-Jensen}, {Dunlop}, {Fynbo}, \& {et al.}}]{Muzzin-13}
{Muzzin}, A., {Marchesini}, D., {Stefanon}, M., {et~al.} 2013, \apj, 777, 18

\bibitem[{{Navarro} {et~al.}(1997){Navarro}, {Frenk}, \& {White}}]{NFW97}
{Navarro}, J.~F., {Frenk}, C.~S., \& {White}, S. D.~M. 1997, \apj, 490, 493

\bibitem[{{Peng} {et~al.}(2015){Peng}, {Maiolino}, \&
  {Cochrane}}]{Peng-Maiolino-Cochrane-15}
{Peng}, Y., {Maiolino}, R., \& {Cochrane}, R. 2015, \nat, 521, 192

\bibitem[{{Peng} {et~al.}(2010){Peng}, {Lilly}, {Kova{\v c}}, {Bolzonella},
  {Pozzetti}, {Renzini}, {Zamorani}, {Ilbert}, \& {et al.}}]{Peng-10}
{Peng}, Y.-j., {Lilly}, S.~J., {Kova{\v c}}, K., {et~al.} 2010, \apj, 721, 193

\bibitem[{{Price-Whelan} {et~al.}(2018){Price-Whelan}, {Sip{\H{o}}cz},
  {G{\"u}nther}, {Lim}, {Crawford}, {Conseil}, {Shupe}, {Craig}, {Dencheva},
  {Ginsburg}, {VanderPlas}, {Bradley}, {P{\'e}rez-Su{\'a}rez}, {de Val-Borro},
  {Paper Contributors}, {Aldcroft}, {Cruz}, {Robitaille}, {Tollerud},
  {Coordination Committee}, {Ardelean}, {Babej}, {Bach}, {Bachetti}, {Bakanov},
  {Bamford}, {Barentsen}, {Barmby}, {Baumbach}, {Berry}, {Biscani}, {Boquien},
  {Bostroem}, {Bouma}, {Brammer}, {Bray}, {Breytenbach}, {Buddelmeijer},
  {Burke}, {Calderone}, {Cano Rodr{\'\i}guez}, {Cara}, {Cardoso}, {Cheedella},
  {Copin}, {Corrales}, {Crichton}, {D{\textquoteright}Avella}, {Deil},
  {Depagne}, {Dietrich}, {Donath}, {Droettboom}, {Earl}, {Erben}, {Fabbro},
  {Ferreira}, {Finethy}, {Fox}, {Garrison}, {Gibbons}, {Goldstein}, {Gommers},
  {Greco}, {Greenfield}, {Groener}, {Grollier}, {Hagen}, {Hirst}, {Homeier},
  {Horton}, {Hosseinzadeh}, {Hu}, {Hunkeler}, {Ivezi{\'c}}, {Jain}, {Jenness},
  {Kanarek}, {Kendrew}, {Kern}, {Kerzendorf}, {Khvalko}, {King}, {Kirkby},
  {Kulkarni}, {Kumar}, {Lee}, {Lenz}, {Littlefair}, {Ma}, {Macleod},
  {Mastropietro}, {McCully}, {Montagnac}, {Morris}, {Mueller}, {Mumford},
  {Muna}, {Murphy}, {Nelson}, {Nguyen}, {Ninan}, {N{\"o}the}, {Ogaz}, {Oh},
  {Parejko}, {Parley}, {Pascual}, {Patil}, {Patil}, {Plunkett}, {Prochaska},
  {Rastogi}, {Reddy Janga}, {Sabater}, {Sakurikar}, {Seifert}, {Sherbert},
  {Sherwood-Taylor}, {Shih}, {Sick}, {Silbiger}, {Singanamalla}, {Singer},
  {Sladen}, {Sooley}, {Sornarajah}, {Streicher}, {Teuben}, {Thomas},
  {Tremblay}, {Turner}, {Terr{\'o}n}, {van Kerkwijk}, {de la Vega}, {Watkins},
  {Weaver}, {Whitmore}, {Woillez}, {Zabalza}, \& {Contributors}}]{astropy-2018}
{Price-Whelan}, A.~M., {Sip{\H{o}}cz}, B.~M., {G{\"u}nther}, H.~M., {et~al.}
  2018, \aj, 156, 123

\bibitem[{{Read} {et~al.}(2006){Read}, {Wilkinson}, {Evans}, {Gilmore}, \&
  {Kleyna}}]{Read-06}
{Read}, J.~I., {Wilkinson}, M.~I., {Evans}, N.~W., {Gilmore}, G., \& {Kleyna},
  J.~T. 2006, \mnras, 366, 429

\bibitem[{{Shi} {et~al.}(2020){Shi}, {Wang}, {Mo}, {Vogelsberger}, {Ho}, {Du},
  {Nelson}, {Pillepich}, \& {Hernquist}}]{Shi2020}
{Shi}, J., {Wang}, H., {Mo}, H., {et~al.} 2020, \apj, 893, 139

\bibitem[{{Simard} {et~al.}(2011){Simard}, {Mendel}, {Patton}, {Ellison}, \&
  {McConnachie}}]{Simard-11}
{Simard}, L., {Mendel}, J.~T., {Patton}, D.~R., {Ellison}, S.~L., \&
  {McConnachie}, A.~W. 2011, \apjs, 196, 11

\bibitem[{{Strateva} {et~al.}(2001){Strateva}, {Ivezi{\'c}}, {Knapp},
  {Narayanan}, {Strauss}, {Gunn}, {Lupton}, {Schlegel}, \& {et
  al.}}]{Strateva-01}
{Strateva}, I., {Ivezi{\'c}}, {\v Z}., {Knapp}, G.~R., {et~al.} 2001, \aj, 122,
  1861

\bibitem[{{Teimoorinia} {et~al.}(2016){Teimoorinia}, {Bluck}, \&
  {Ellison}}]{Teimoorinia-Bluck-Ellison-16}
{Teimoorinia}, H., {Bluck}, A.~F.~L., \& {Ellison}, S.~L. 2016, \mnras, 457,
  2086

\bibitem[{{Tomczak} {et~al.}(2014){Tomczak}, {Quadri}, {Tran}, {Labb{\'e}},
  {Straatman}, {Papovich}, {Glazebrook}, {Allen}, \& {et al.}}]{Tomczak-14}
{Tomczak}, A.~R., {Quadri}, R.~F., {Tran}, K.-V.~H., {et~al.} 2014, \apj, 783,
  85

\bibitem[{{Toomre} \& {Toomre}(1972)}]{Toomre-Toomre-72}
{Toomre}, A., \& {Toomre}, J. 1972, \apj, 178, 623

\bibitem[{{van den Bosch} {et~al.}(2008){van den Bosch}, {Aquino}, {Yang},
  {Mo}, {Pasquali}, {McIntosh}, {Weinmann}, \& {Kang}}]{vandenBosch-08}
{van den Bosch}, F.~C., {Aquino}, D., {Yang}, X., {et~al.} 2008, \mnras, 387,
  79

\bibitem[{{van der Walt} {et~al.}(2011){van der Walt}, {Colbert}, \&
  {Varoquaux}}]{numpy}
{van der Walt}, S., {Colbert}, S.~C., \& {Varoquaux}, G. 2011, Computing in
  Science Engineering, 13, 22

\bibitem[{{van der Wel} {et~al.}(2014){van der Wel}, {Franx}, {van Dokkum},
  {Skelton}, {Momcheva}, {Whitaker}, {Brammer}, {Bell}, \& {et
  al.}}]{vanderWel-14}
{van der Wel}, A., {Franx}, M., {van Dokkum}, P.~G., {et~al.} 2014, \apj, 788,
  28

\bibitem[{{Wang} {et~al.}(2018{\natexlab{a}}){Wang}, {Wang}, {Mo}, {van den
  Bosch}, {Lim}, {Wang}, {Yang}, \& {Chen}}]{WangE-18b}
{Wang}, E., {Wang}, H., {Mo}, H., {et~al.} 2018{\natexlab{a}}, \apj, 864, 51

\bibitem[{{Wang} {et~al.}(2020){Wang}, {Wang}, {Mo}, {van den Bosch}, \&
  {Yang}}]{WangE-20}
{Wang}, E., {Wang}, H., {Mo}, H., {van den Bosch}, F.~C., \& {Yang}, X. 2020,
  \apj, 889, 37

\bibitem[{{Wang} {et~al.}(2015){Wang}, {Wang}, {Kauffmann}, {J{\'o}zsa}, \&
  {Li}}]{WangE-15}
{Wang}, E., {Wang}, J., {Kauffmann}, G., {J{\'o}zsa}, G.~I.~G., \& {Li}, C.
  2015, \mnras, 449, 2010

\bibitem[{{Wang} {et~al.}(2018{\natexlab{b}}){Wang}, {Wang}, {Mo}, {Lim}, {van
  den Bosch}, {Kong}, {Wang}, {Yang}, \& {et al.}}]{WangE-18a}
{Wang}, E., {Wang}, H., {Mo}, H., {et~al.} 2018{\natexlab{b}}, \apj, 860, 102

\bibitem[{{Wang} {et~al.}(2012){Wang}, {Mo}, {Yang}, \& {van den
  Bosch}}]{WangH-12}
{Wang}, H., {Mo}, H.~J., {Yang}, X., \& {van den Bosch}, F.~C. 2012, \mnras,
  420, 1809

\bibitem[{{Wang} {et~al.}(2016){Wang}, {Mo}, {Yang}, {Zhang}, {Shi}, {Jing},
  {Liu}, {Li}, \& {et al.}}]{WangH-16}
{Wang}, H., {Mo}, H.~J., {Yang}, X., {et~al.} 2016, \apj, 831, 164

\bibitem[{{Wang} {et~al.}(2018{\natexlab{c}}){Wang}, {Mo}, {Chen}, {Yang},
  {Yang}, {Wang}, {van den Bosch}, {Jing}, \& {et al.}}]{WangH-18}
{Wang}, H., {Mo}, H.~J., {Chen}, S., {et~al.} 2018{\natexlab{c}}, \apj, 852, 31

\bibitem[{{Weinmann} {et~al.}(2009){Weinmann}, {Kauffmann}, {van den Bosch},
  {Pasquali}, {McIntosh}, {Mo}, {Yang}, \& {Guo}}]{Weinmann-09}
{Weinmann}, S.~M., {Kauffmann}, G., {van den Bosch}, F.~C., {et~al.} 2009,
  \mnras, 394, 1213

\bibitem[{{Weinmann} {et~al.}(2006){Weinmann}, {van den Bosch}, {Yang}, \&
  {Mo}}]{Weinmann-06}
{Weinmann}, S.~M., {van den Bosch}, F.~C., {Yang}, X., \& {Mo}, H.~J. 2006,
  \mnras, 366, 2

\bibitem[{{Wetzel} {et~al.}(2012){Wetzel}, {Tinker}, \&
  {Conroy}}]{Wetzel-Tinker-Conroy-12}
{Wetzel}, A.~R., {Tinker}, J.~L., \& {Conroy}, C. 2012, \mnras, 424, 232

\bibitem[{{White} \& {Frenk}(1991)}]{White-Frenk-91}
{White}, S.~D.~M., \& {Frenk}, C.~S. 1991, \apj, 379, 52

\bibitem[{{White} \& {Rees}(1978)}]{White-Rees-78}
{White}, S.~D.~M., \& {Rees}, M.~J. 1978, \mnras, 183, 341

\bibitem[{{Woo} {et~al.}(2015){Woo}, {Dekel}, {Faber}, \& {Koo}}]{Woo-15}
{Woo}, J., {Dekel}, A., {Faber}, S.~M., \& {Koo}, D.~C. 2015, \mnras, 448, 237

\bibitem[{{Woo} {et~al.}(2013){Woo}, {Dekel}, {Faber}, {Noeske}, {Koo},
  {Gerke}, {Cooper}, {Salim}, \& {et al.}}]{Woo-13}
{Woo}, J., {Dekel}, A., {Faber}, S.~M., {et~al.} 2013, \mnras, 428, 3306

\bibitem[{{Yang} {et~al.}(2009){Yang}, {Mo}, \& {van den
  Bosch}}]{Yang-Mo-vandenBosch-09}
{Yang}, X., {Mo}, H.~J., \& {van den Bosch}, F.~C. 2009, \apj, 695, 900

\bibitem[{{Yang} {et~al.}(2007){Yang}, {Mo}, {van den Bosch}, {Pasquali}, {Li},
  \& {Barden}}]{Yang-07}
{Yang}, X., {Mo}, H.~J., {van den Bosch}, F.~C., {et~al.} 2007, \apj, 671, 153

\bibitem[{{Yang} {et~al.}(2005){Yang}, {Mo}, {van den Bosch}, {Weinmann}, {Li},
  \& {Jing}}]{Yang2005b}
---. 2005, \mnras, 362, 711

\end{thebibliography}

\appendix

%\section{Tests narrower redshift ranges.}

%In appendix, we present two results. First is on the SFR-$M_*$ diagram in different $B/T$, second is about the impact of the observational effect and performance of some techniques used to derive the quantities used in the paper.

\begin{figure}
    \centering
    \includegraphics[scale=0.5]{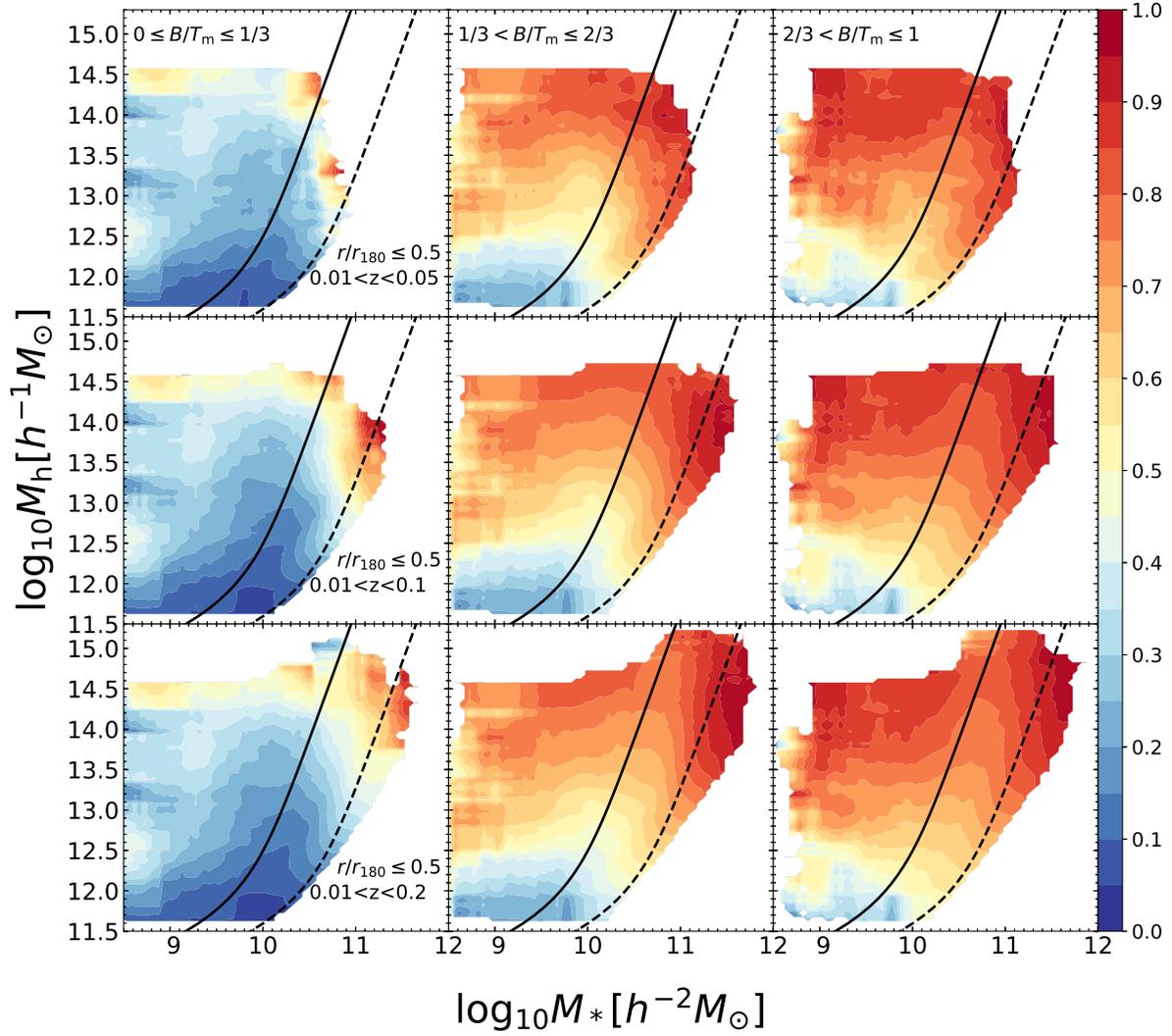}
	\caption{Contours show $f_{\rm Q}$ as a function of $M_{\rm h}$
	and $M_{*}$ for galaxies with $r/r_{180} \leq 0.5$. 
	The top, middle and bottom panels show results for galaxies with $0.01<z<0.05$, $0.01<z<0.1$ and $0.01<z<0.2$, respectively.
	Different columns show different $B/T_{\rm m}$, as indicated in the top panels. 
	The results shown are smoothed on a grid with cell size 
	given by $\Delta \log(M_*/\msunt)=0.6$ and 
	$\Delta \log(M_{\rm h}/\msun)=0.6$. Cells with less than 
	10 galaxies are discarded. The black solid line shows 
	Equation (\ref{eq_mc}) and the dashed line shows the central mass 
	and halo mass relation given by \cite{Yang-Mo-vandenBosch-09}.}
	\label{fq_Mh_Mstellar_BTm_r_contour_z}\label{fig_lowz}
\end{figure}

Sample completeness, fiber aperture and the uncertainties in  
bulge$+$disk decomposition may generate uncertainties in the physical 
quantities used in our analysis. These effects are usually redshift-dependent.
For example, fiber aperture effect is expected to be more significant 
at lower $z$ while bulge$+$disk decomposition and group/galaxy 
samples may be more uncertain at higher $z$. 
In addition, galaxy groups with $\log(M_{\rm h}/\msun)=11.6$ 
are only complete at $z<0.10$, although more massive groups are complete 
to higher $z$ \citep{Yang-Mo-vandenBosch-09,WangH-12}.
To examine the impact of such effects, we have made tests using 
galaxies in two narrower redshift ranges, 
$0.01<z\le 0.05$ and $0.01<z\le 0.1$.
Since the sample size is reduced by using galaxies in a narrower redshift 
range and since we are considering 
the quenched fraction in a multi-dimensional space, 
we have to use larger $B/T_{\rm m}$ bins and larger 
smoothing length to suppress noise. 
The results are shown in Figure \ref{fig_lowz}. 
For comparison, the results for the whole sample 
(i.e. $0.01<z<0.2$) analyzed in the same way are shown in 
the bottom panels. 
As one can see, the transition feature is clearly seen in all 
the three $B/T_{\rm m}$ bins for the low-$z$ samples.
More importantly, the behavior of the transition
in the $(M_*, M_{\rm h})$ space is very similar to that 
for the whole sample. The transition feature for the lowest 
redshift sample appears to 
be slightly weaker than that for the other two samples. 
There are two reasons for this. Firstly, as mentioned above, 
since the size of this sample is much smaller, 
the statistic is noisier, 
particularly for galaxies and groups of high masses whose 
number densities are low. Indeed, the contours at the upper-right 
corner are now missing exactly because of the lack of galaxies 
in this region. Secondly, fiber aperture effect may affect the 
measurements of star formation rates, although corrections 
were made for aperture bias in the original data 
\citep{Brinchmann-04}. The aperture effect is expected to be severer at 
lower redshift, because of the larger angular sizes of
galaxies. We also checked the quenched fraction as a function 
of $(r/r_{180}, M_*)$ using the two lower-$z$ samples, and 
found that all of our conclusions remain unchanged.
These test results demonstrate that the transition feature and 
the characteristic mass are not generated by these effects.

\begin{figure*}
  \centering
    \includegraphics[scale=0.5]{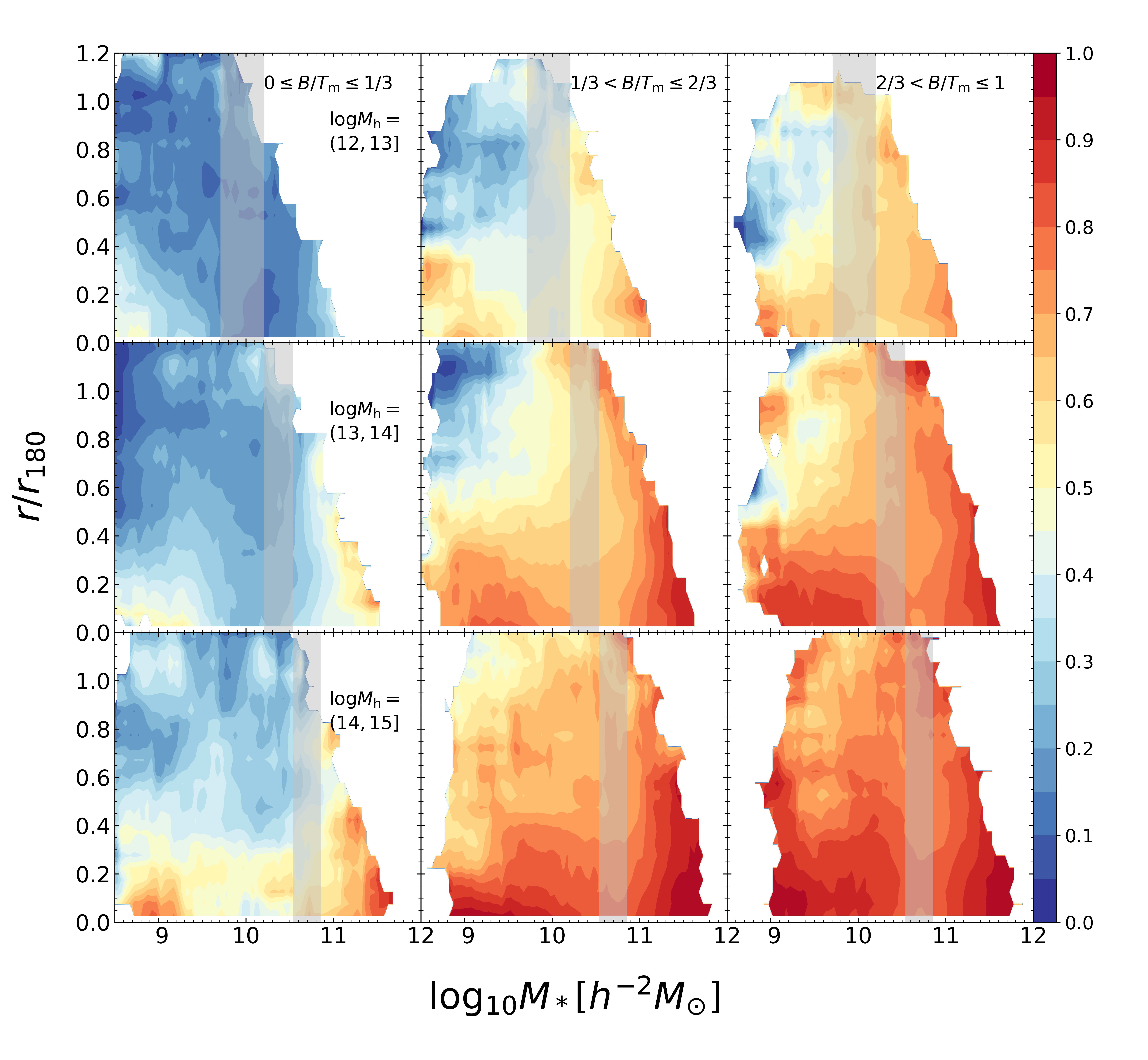}
	\caption{Contours show $f_{\rm Q}$ as a function of $r/r_{180}$ and $M_*$. 
	Different rows correspond to different $M_{\rm h}$ and different 
	columns correspond to different $B/T_{\rm m}$. The grey area in each panel 
	shows the $M_{\rm *, ch}$ range corresponding to the $M_{\rm h}$ range. 
	The grid cell size used in the plot is given by 
	$\Delta \log(M_*/\msunt)=0.6$ and $\Delta r/r_{180}=0.2$;
	the results are insensitive to changes in the cell size.  
	Cells containing less than 10 galaxies are discarded. 
	}\label{fq_r_Mstellar_BTm_Mh_contour}
\end{figure*}

The results shown in Figure \ref{fq_r_Mstellar_BTm_Mtstar_contour}
do not distinguish halos of different mass.  
To examine whether or not the dependence on $r/r_{180}$ shown 
there is caused by the dependence on halo mass, 
we show in Figure \ref{fq_r_Mstellar_BTm_Mh_contour}
the quenched fraction as a function of $r/r_{180}$ and $M_*$ in 
three $M_{\rm h}$ bins.  We use only three $B/T_{\rm m}$ bins 
in order to reduce statistical uncertainties. 
The vertical band in each panel indicates the range of 
$M_{\rm *,ch}$ covered by the corresponding halo mass 
bin.  As one can see, even within a given halo mass bin, 
galaxies below and above $M_{\rm *, ch}$ have different
dependence on $r/r_{180}$. The dependence is strong 
at $M_*< M_{\rm *, ch}$, but rather weak at 
$M_*> M_{\rm *, ch}$, indicating that the $r/r_{180}$-dependence 
within each class shown in Figure \ref{fq_r_Mstellar_BTm_Mtstar_contour} 
is not caused by the dependence on halo mass.

\label{lastpage}
\end{document}